\documentclass[aps,pra,reprint,amsfonts,amssymb,amsmath,longbibliography,superscriptaddress]{revtex4-2}

\usepackage{amssymb,amsmath,bm}
\usepackage{fullpage}
\usepackage{graphicx}
\usepackage{wrapfig}
\usepackage{floatflt}
\usepackage{color}
\usepackage{xspace}
\usepackage{dsfont}
\usepackage[official,right]{eurosym}
\usepackage[top=0.8in, bottom=0.8in, left=0.65in, right=0.65in]{geometry}
\usepackage[utf8]{inputenc}
\usepackage{hyperref}

\newcommand{\CsFe}{CsFeCl$_3$\xspace}
\newcommand{\RbFe}{RbFeCl$_3$\xspace}

\newcommand{\be}{\begin{equation} }
\newcommand{\ee}{\end{equation} }
\newcommand{\bea}{\begin{eqnarray} }
\newcommand{\eea}{\end{eqnarray} }

\begin{document}
\title{Anomalous spin waves in CsFeCl$_{3}$ and RbFeCl$_{3}$}

\author{L.~Stoppel}
\affiliation{Laboratory for Solid State Physics, ETH Z{\"u}rich, 8093 Z{\"u}rich, Switzerland}
\author{S.~Hayashida}
\email[Corresponding author: ]{shoheih@phys.ethz.ch}
\affiliation{Laboratory for Solid State Physics, ETH Z{\"u}rich, 8093 Z{\"u}rich, Switzerland}
\author{Z.~Yan}
\affiliation{Laboratory for Solid State Physics, ETH Z{\"u}rich, 8093 Z{\"u}rich, Switzerland}
\author{A.~Podlesnyak}
\affiliation{Neutron Scattering Division, Oak Ridge National Laboratory, Oak Ridge, Tennessee 37831, USA}
\author{A. Zheludev}
\email{zhelud@ethz.ch}
\homepage{http://www.neutron.ethz.ch/}
\affiliation{Laboratory for Solid State Physics, ETH Z{\"u}rich, 8093 Z{\"u}rich, Switzerland}

\date{\today}

\begin{abstract}
We investigate magnetic excitations in the $S=1$ easy-plane-type triangular antiferromagnets \CsFe and \RbFe through a combination of inelastic neutron scattering measurements  and spin-wave theory calculations based on an established exchange-coupling Hamiltonian. We show that in either material the model falls short of providing an adequate description of the measured intensities and for \RbFe even fails to reproduce the measured dispersion relation. The most striking discrepancy  is a very anisotropic azimuthal intensity distribution in the acoustic spin-wave branches in the long-wavelength limit, which is incompatible with spin-wave theory on a fundamental level. The observed anomalies are attributed to long-range dipolar interactions.
\end{abstract}

\maketitle

\section{Introduction}\label{sec1}
The alkali-metal trichloroferrates $A$FeCl$_{3}$ ($A=$ Cs and Rb) are prototypical  $S=1$ triangular lattice antiferromagnets with strong planar magnetic anisotropy~\cite{Montano1973,Montano1974,Eibschutz1975,Yoshizawa1980,Haseda1981}.
Under the influence of hydrostatic pressure~\cite{Kurita2016,Hayashida2018,Hayashida2019pressure} or chemical composition~\cite{Hayashida2019chemical} they are known to exhibit quantum phase transitions between gapped quantum-paramagnetic and magnetically ordered states.
Spin-wave excitations in both compounds have been extensively studied and interpreted based on a model Hamiltonian with short-range interactions~\cite{Yoshizawa1980,Petitgrand1981,Suzuki1983,Hayashida2019pressure}.
The latter includes easy-plane single-ion anisotropy, ferromagnetic chains, and antiferromagnetic coupling in the triangular lattice that the chains are arranged in.
At the same time, it is well established that long-ranged magnetic dipolar interactions may also be relevant.
In particular, these are seen as being responsible for additional narrow incommensurate magnetic phases found in \RbFe  between the
classic 120$^\circ$ structure at low temperature and the high-temperature paramagnetic phases ~\cite{Wada1982,Shiba1982,Shiba1983,Suzuki1983dipole}.
What effect, if any, do these interactions have on the spin-wave excitations in these species?

To answer this question, we undertake a detailed quantitative analysis of the inelastic neutron scattering (INS) experiments briefly reported in our previous paper~\cite{Hayashida2019chemical}. We utilize the so-called extended spin-wave theory (ESWT) ~\cite{Shiina2003,Matsumoto2008} based on a Hamiltonian with short-range interactions.
For \CsFe, this approach seems to work reasonably well but requires the introduction of additional exchange constants not considered in the previous studies~\cite{Yoshizawa1980,Petitgrand1981,Suzuki1983,Hayashida2019pressure}. 
Even so, some quantitative discrepancies between experiment and calculation remain.
For \RbFe though, ESWT fails already on a qualitative level. 
The most striking finding is that in the long-wavelength limit the measured spin-wave intensity is
highly anisotropic near magnetic Bragg peaks. In contrast, spin-wave analysis based on short-range interactions alone predicts an isotropic intensity distribution.
Other important discrepancies are apparent in certain features of the dispersion curves.

\section{Material and experiment}\label{sec2}
\begin{figure}[tbp]
\includegraphics[scale=1]{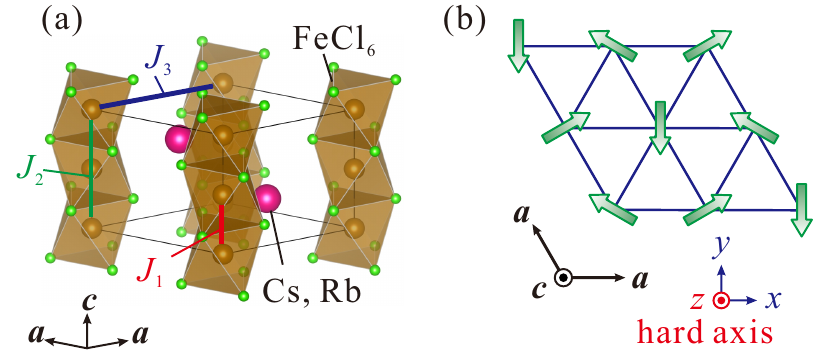}
\caption{(a) The crystal structure of $A$FeCl$_{3}$ ($A=$ Cs, Rb) with the space group $P6_{3}/mmc$.
Solid lines are exchange paths of the intrachain nearest neighbor ($J_{1}$) and next-nearest neighbor ($J_{2}$) and the in-plane nearest neighbor ($J_{3}$).
(b) The 120$^{\circ}$ spin structure in the triangular lattice and the geometry conventions used in this paper.}
\label{fig:crystal}
\end{figure}

Both \CsFe and \RbFe crystallize in a hexagonal structure with the space group $P6_{3}/mmc$~\cite{Kohne1993}.
The lattice parameters are as reported in Ref.~\cite{Hayashida2019chemical}.
Magnetism is due to Fe$^{2+}$ ions ($3d^{6}$, $S=2$, $L=2$).
Face-sharing FeCl$_{6}$ octahedra form one-dimensional chains along the crystallographic $c$ axis.
The chains form a triangular lattice in the $ab$ plane [Fig.~\ref{fig:crystal}(a)].
The low-energy excitation of the Fe$^{2+}$ ion is described by a pseudospin $s=1$ because of the cubic crystal-field and spin-orbit coupling~\cite{Eibschutz1975}.
In the following, we represent the pseudospin as $S=1$ for convenience.
\CsFe is a quantum paramagnet, with a gapped disordered ground state~\cite{Yoshizawa1980,Haseda1981}.
\RbFe shows magnetic long-range order (LRO) below 2.6~K~\cite{Haseda1981}.
Below 1.95~K the structure is a 120$^{\circ}$ spin arrangement with a commensurate propagation vector $\mathbf{Q}=(1/3,1/3,0)$~\cite{Wada1982}, as displayed in Fig.~\ref{fig:crystal}(b).

Details of the neutron scattering experiment to measure the spin-wave spectrum were reported in Ref.~\cite{Hayashida2019chemical}.
The data were collected at the Cold Neutron Chopper Spectrometer (CNCS)~\cite{cncs2011,cncs2016} installed at the Spallation Neutron Source (SNS) of Oak Ridge National Laboratory, Oak Ridge, Tennessee.
The energies of the incident neutron beam were set at $E_{\rm i}=5.93$ and 2.99~meV, yielding energy resolutions of $\Delta E=0.23$ and 0.09 meV at the elastic position, respectively.
All data were analyzed using HORACE software~\cite{horace2016}

\begin{figure*}[tbp]
\includegraphics[scale=1]{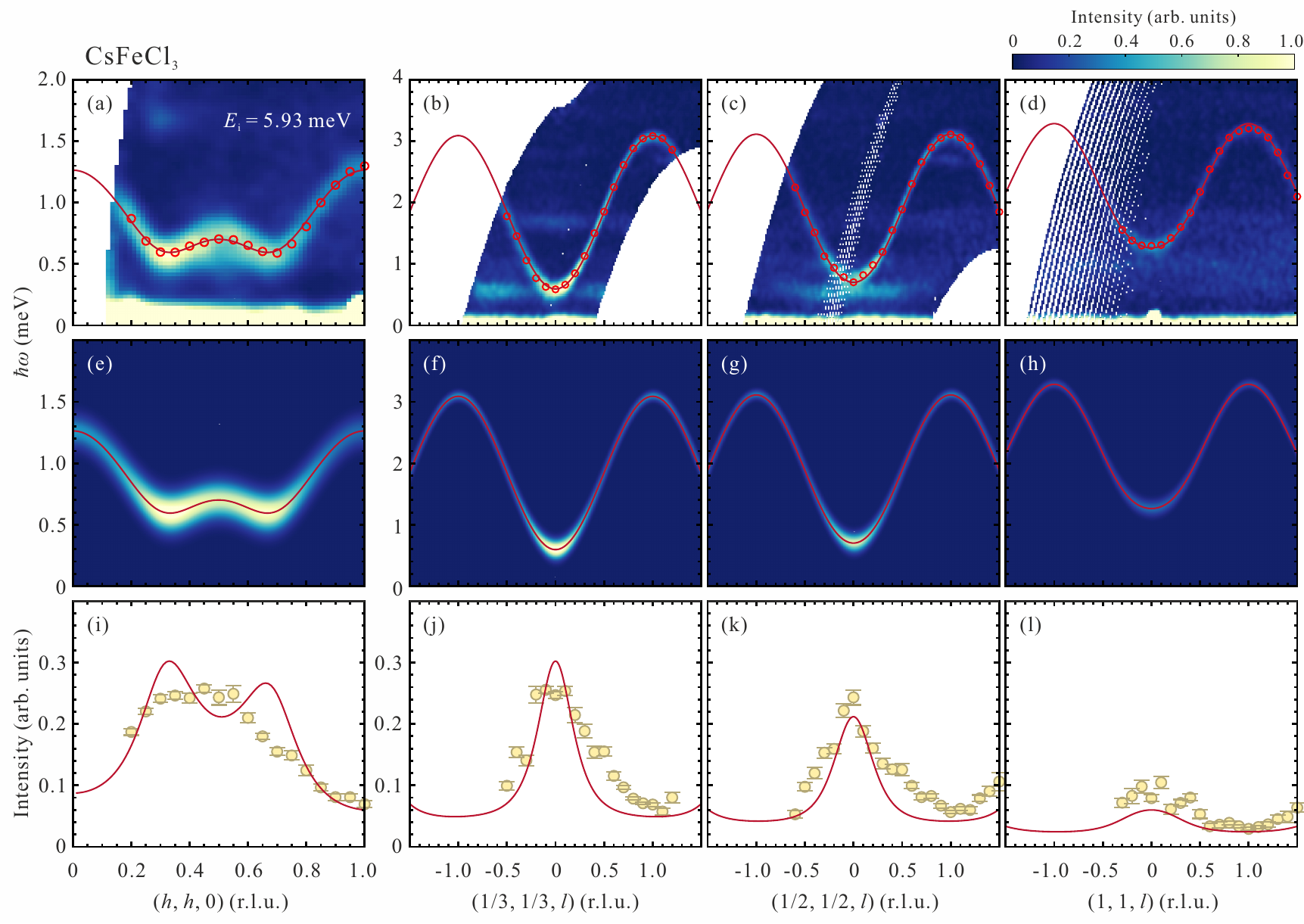}
\caption{(a)--(d) False-color plots of inelastic neutron scattering intensities measured at  $T\simeq 0.1$~K in \CsFe.
(a) and (b) are the same as plots given in Ref.~\cite{Hayashida2019chemical}.
All intensities are integrated in momentum transfer perpendicular to the plot axis in the range of $|q|\leq 0.05$~{\AA}$^{-1}$.
Red open circles are the peak positions extracted by the fits to individual constant-$\mathbf{q}$ cuts. The fit variances are inside the symbols.
(e)--(h) Simulated ESWT spectra with the background intensity of 0.03, convoluted with the calculated instrumental resolution at $E_{\rm i}=5.93$~meV.
In all cases, solid curves represent the fitted calculated dispersion.
(i)--(l) Integrated intensities determined in fits to individual constant-$\mathbf{q}$ cuts (symbols) and the ESWT calculation based on Hamiltonian parameters determined in a global fit to the measured dispersion relation (solid curve, arbitrary scaling that is uniform across the four panels).
}
\label{fig:INS_Cs}
\end{figure*}
\begin{figure*}[tbp]
\includegraphics[scale=1]{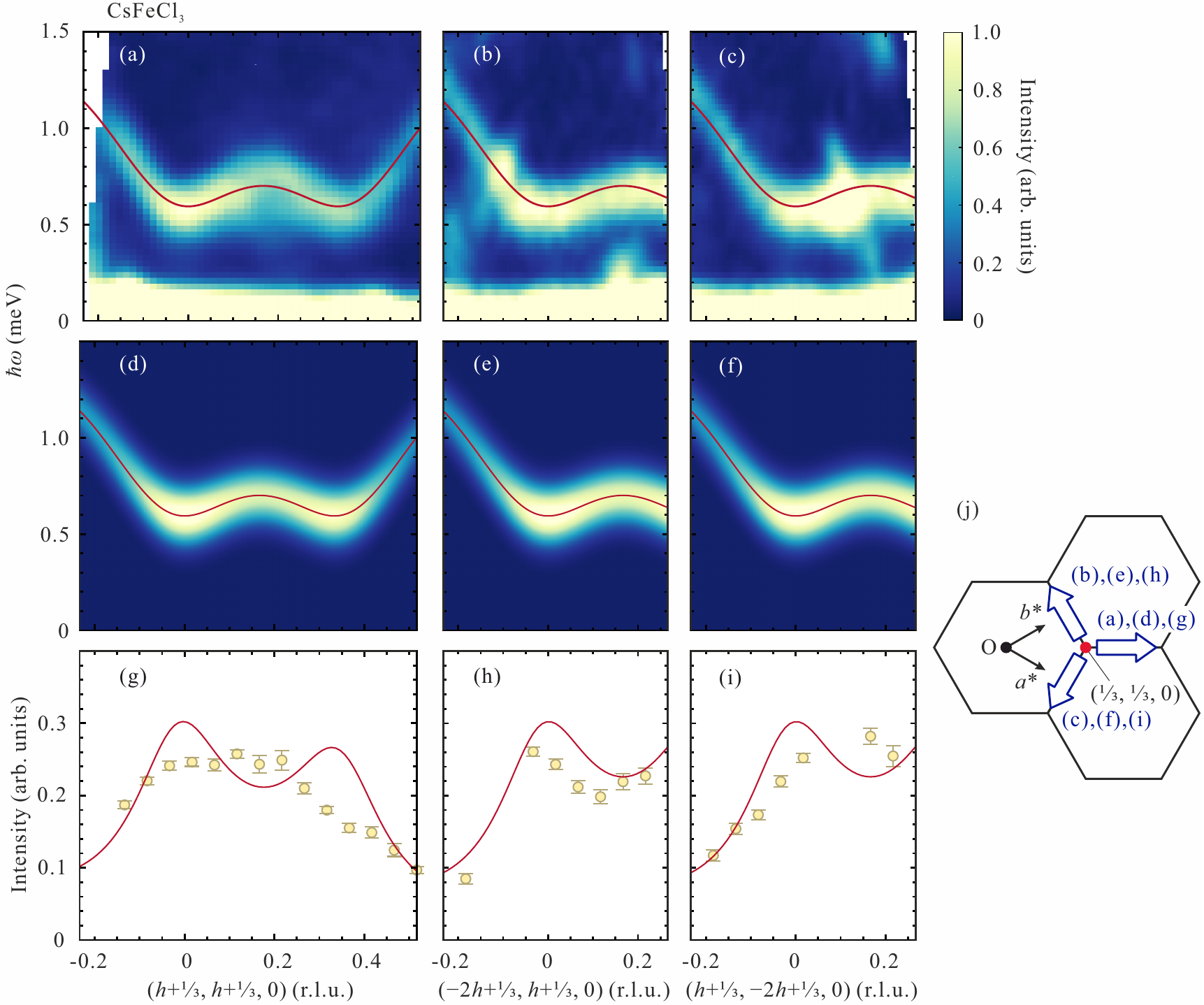}
\caption{(a)--(i) Similar to Fig.~\ref{fig:INS_Cs} for different equivalent scattering directions in the triangular plane.  The scale factor of the ESWT calculation in (g)--(i) is arbitrary and uniform across the three panels. 
Integrated intensity data affected by the spurious bands are not shown.
(j) Orientation of the shown energy-momentum slices in reciprocal space. The black hexagon are crystallographic Brillouin zones.}
\label{fig:threeQ_Cs}
\end{figure*}

\section{Theoretical approach}\label{sec3}
To quantitatively analyze the spin-wave excitations, we employ the ESWT approach~\cite{Shiina2003,Matsumoto2008}.
It is equivalent to the harmonic bond-operator theory~\cite{Sachdev1990,Sommer2001,Matsumoto2002,Matsumoto2004} and is applied to noncollinear spin structure systems~\cite{Matsumoto2014}.
We consider an effective Hamiltonian of the $S=1$ easy-plane-type antiferromagnet as follows:
\begin{eqnarray}
{\cal H}&=&J_{1}\sum_{\rm n.n.}^{\rm chain}\mathbf{S}_{i}\cdot\mathbf{S}_{j}+J_{2}\sum_{\rm n.n.n.}^{\rm chain}\mathbf{S}_{i}\cdot\mathbf{S}_{j} \nonumber \\
&&+J_{3}\sum_{\rm n.n.}^{\rm plane}\mathbf{S}_{i}\cdot\mathbf{S}_{j}+D\sum_{i}^{\rm site}(S^{z})^{2}, \label{eq:Hamiltonian}
\end{eqnarray}
where $J_{1}$ and $J_{2}$ are the nearest- and next-nearest-neighbor interactions in the chain and $J_{3}$ is the nearest-neighbor in-plane interaction.
The exchange paths are displayed in Fig.~\ref{fig:crystal}(a).
The summation is taken over all sites connected by $J_{1}$, $J_{2}$, and $J_{3}$.
Note that the possibility of a $J_2$ coupling has not been considered in previous studies~\cite{Yoshizawa1980,Petitgrand1981,Suzuki1983,Hayashida2019pressure}.
$D$ is positive and gives the easy-plane single-ion anisotropy.
This anisotropy splits the triplet spin $S=1$ state into the singlet $S^{z}=0$ and the doublet $S^{z}=\pm 1$ states, thus supporting a gapped paramagnetic state.
In contrast, the spin interactions favor magnetic order.

The ESWT calculation on INS spectra for Hamiltonian~(\ref{eq:Hamiltonian}) is reported in Refs.~\cite{Hayashida2019pressure,Matsumoto2020}.  The approach is applicable to both the gapped and ordered phases. The spectra in the two regimes can be described as follows.
In the gapped  phase (as in \CsFe), two degenerated gapped modes $\hbar\omega(\mathbf{q})$ emerge.
In the ordered state (as in \RbFe), the degenerated modes split into gapless $\hbar\omega_{\rm T}$ and gapped $\hbar\omega_{\rm L}$ modes.
Owing to the three sublattices of the $120^{\circ}$ structure, six modes are generated in total: $\hbar\omega_{\rm T}(\mathbf{q})$, $\hbar\omega_{\rm L}(\mathbf{q})$, $\hbar\omega_{\rm T}(\mathbf{q}\pm\mathbf{Q})$, and $\hbar\omega_{\rm L}(\mathbf{q}\pm\mathbf{Q})$.
The gapless $\hbar\omega_{\rm T}$ and gapped $\hbar\omega_{\rm L}$ modes are mainly associated with the transverse and longitudinal fluctuations of the spins, respectively, even though they are hybridized to some extent due to the noncollinear spin structure~\cite{Hayashida2019pressure,Matsumoto2020}.

The dynamical structure factor (intensity) $S^{zz}(\mathbf{q},\omega)$ for $\hbar\omega_{\rm T}(\mathbf{q})$ and $\hbar\omega_{\rm L}(\mathbf{q})$ is much smaller than  $S^{xx}(\mathbf{q},\omega)$ for $\hbar\omega_{\rm T}(\mathbf{q}\pm\mathbf{Q})$ and $\hbar\omega_{\rm L}(\mathbf{q}\pm\mathbf{Q})$ because the former represents spin fluctuations along the hard axis [see the coordinates in the schematics of Fig.~\ref{fig:crystal}(b)].
In the following, we thus display only $\hbar\omega_{\rm T}(\mathbf{q}\pm\mathbf{Q})$ and $\hbar\omega_{\rm L}(\mathbf{q}\pm\mathbf{Q})$.
For plotting $S^{xx}(\mathbf{q},\omega)$, polarization factors and the magnetic form factor are taken into account as reported in Refs.~\cite{Hayashida2019pressure,Matsumoto2020}.

\begin{figure*}[tbp]
\includegraphics[scale=1]{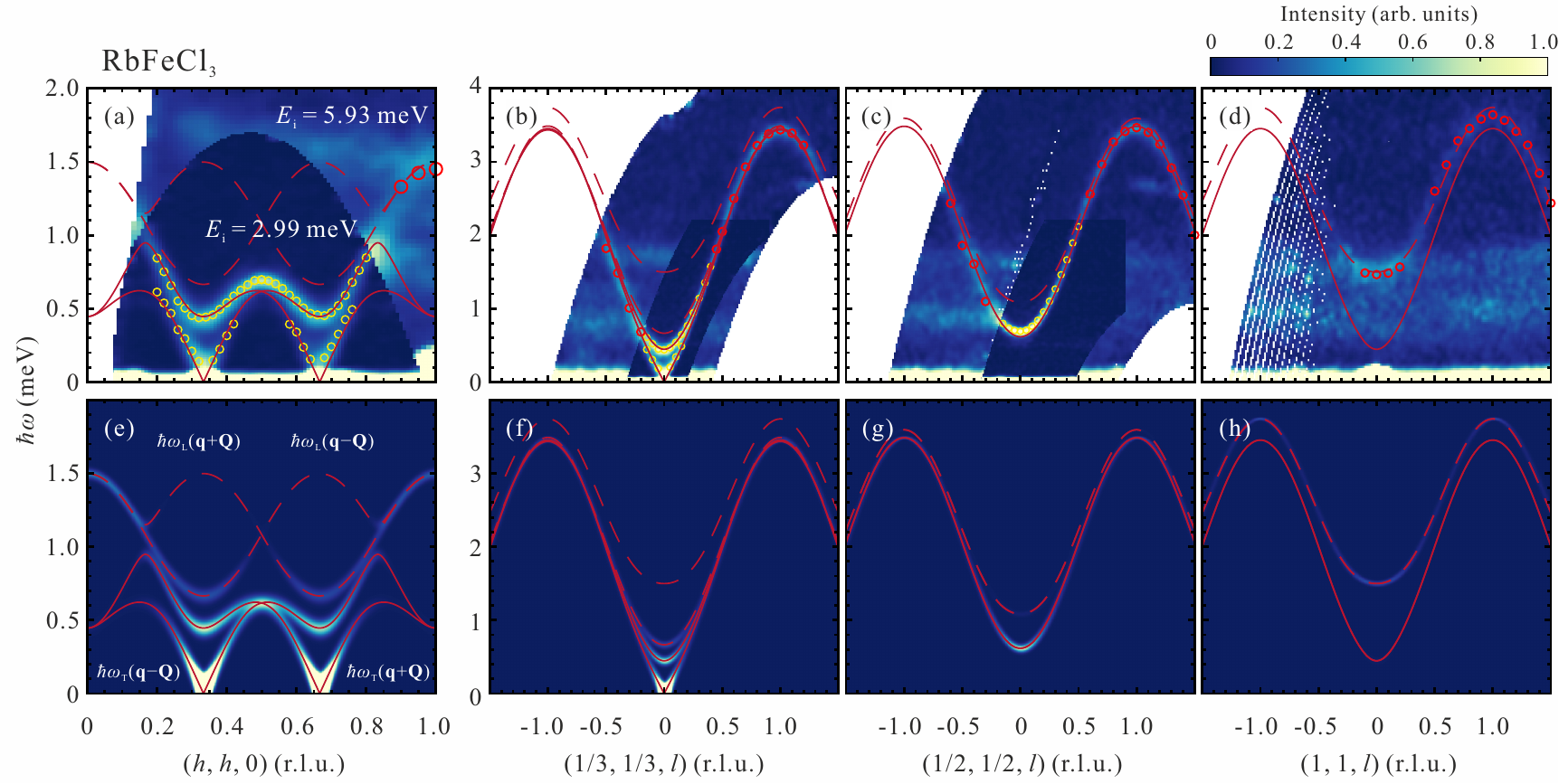}
\caption{(a)--(d) False-color plots of the INS spectra at $T\simeq 0.1$~K for \RbFe. The spectra with $E_{i}=2.99$~meV are plotted over the spectra with $E_{i}=5.93$~meV.
(a) and (b) are the same as plots given in Ref.~\cite{Hayashida2019chemical}.
All intensities are integrated in momentum transfer perpendicular to the plot axis in the range of $|q|\leq 0.05$~{\AA}$^{-1}$.
Yellow and red open circles are the peak positions extracted by the fits to individual constant-$\mathbf{q}$ cuts for $E_{i}=2.99$ and 5.93~meV, respectively. The error bars are inside the symbols.
(e)--(h) Simulated ESWT spectra with the background intensity of 0.01, convoluted with the instrumental resolution at $E_{\rm i}=2.99$~meV.
The solid curves in all the plots are the simulated dispersion.}
\label{fig:INS_Rb}
\end{figure*}

\begin{figure*}[tbp]
\includegraphics[scale=1]{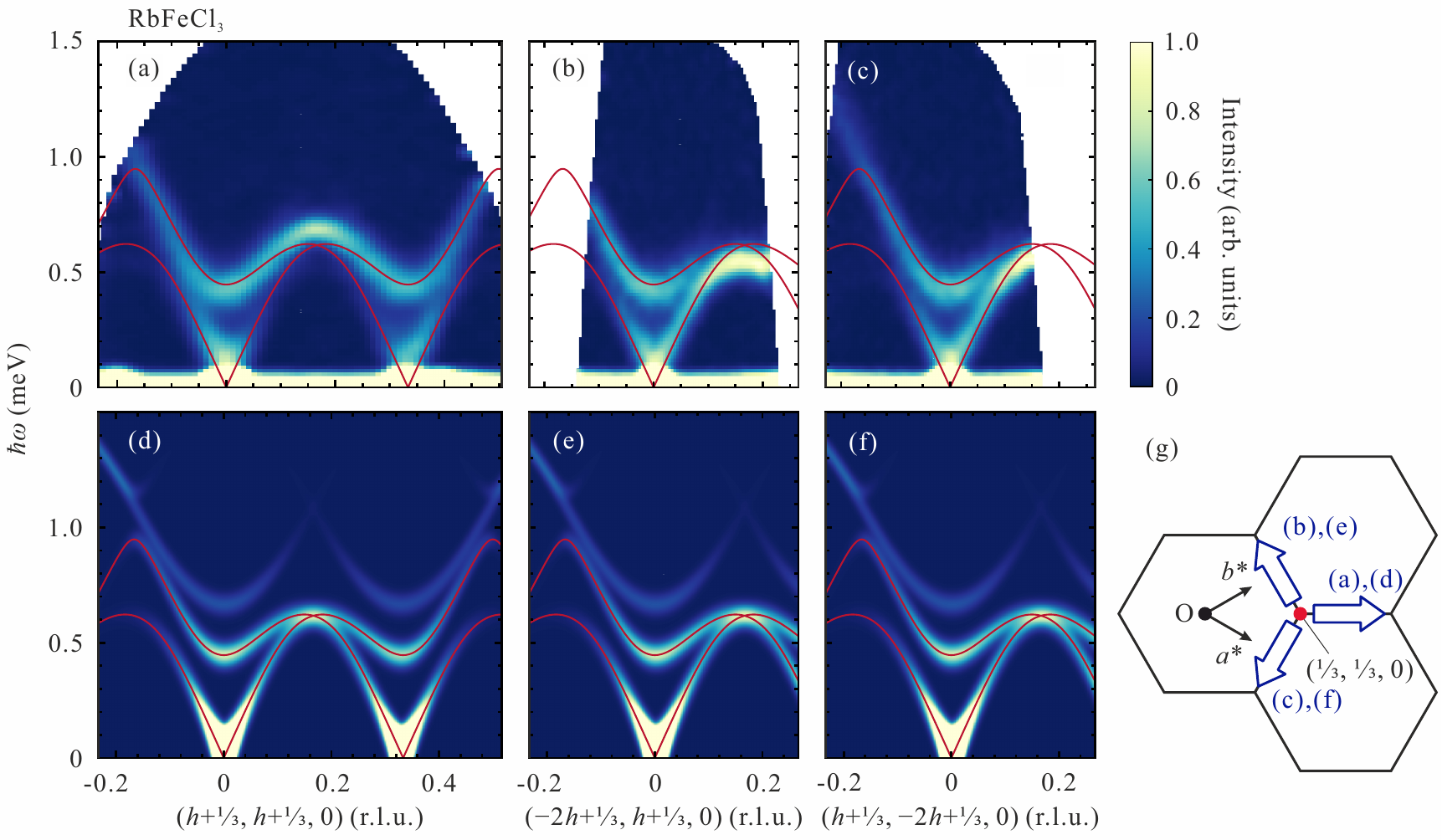}
\caption{INS and simulated spectra projected onto the (a) and (d) $\hbar\omega$-$(h,h,0)$, (b) and (e) $\hbar\omega$-$(-2h+1/3,h+1/3,0)$, and (c) and (f) $\hbar\omega$-$(h+1/3,-2h+1/3,0)$ planes in \RbFe.
(g) A schematic diagram of the reciprocal space in the $a^{*}b^{*}$ plane. The black hexagons show crystallographic Brillouin zones. The blue arrows are the projected directions for (a)--(f).}
\label{fig:threeQ_Rb}
\end{figure*}

\section{Results}
\subsection{\CsFe: quantum paramagnetic state}\label{sec:CsFeCl3}
For \CsFe, a gapped dispersion relation is observed below 4~meV, and the ESWT calculation is in excellent agreement with the observed dispersion [Figs.~\ref{fig:INS_Cs}(a)--\ref{fig:INS_Cs}(d)].
We quantified the measured dispersion by analyzing individual constant-$\mathbf{q}$ cuts of the data, in which peaks at each wave vector were fitted by the Gaussian function.
Note that the flat intensity bands at $\hbar\omega=0.8$ and 1.8 meV are spurious as mentioned in Ref.~\cite{Hayashida2019chemical}.
For data in the $ab$ plane, the fitted peak linewidths vary between the computed energy resolution of the instrument (approximately 0.2~meV at a typical 0.5-meV energy transfer for $E_i=5.93$~meV) at $(1/3,1/3,0)$ and almost twice that value at $(1/2,1/2,0)$.
The Hamiltonian parameters are evaluated from a global least-squares fitting of the dispersion relation calculated along the four directions to the one measured.
The best agreement is obtained by allowing for a weak antiferromagnetic next-nearest-neighbor interaction $J_{2}$ in the chains, which was disregarded in previous studies~\cite{Yoshizawa1980,Petitgrand1981,Suzuki1983,Hayashida2019pressure}.
The fitted parameters and variances are listed in Table~\ref{tb:parameters} and correspond to a fit with $\chi^{2}=1.3$.

At first glance, the neutron scattering intensities simulated using the parameters obtained from analyzing the dispersion agree well with the measured spectra, as shown in Figs.~\ref{fig:INS_Cs}(e)--\ref{fig:INS_Cs}(h). 
These false-color plots are obtained by convolution of the ESWT-computed intensities with the calculated energy-dependent resolution of the spectrometer.
A more careful look reveals quantitative discrepancies. Figures~\ref{fig:INS_Cs}(i)--\ref{fig:INS_Cs}(l) show the measured energy-integrated intensities (symbols) in comparison with the ESWT result. The scaling of the ESWT curves is arbitrary but consistent between the four panels. One immediately notices that the measured intensities seem to lack distinct maxima predicted near the zone centers of the 120$^\circ$-structure $\mathbf{q}=(1/3,1/3,0)$ and $(2/3,2/3,0)$.

Figures~\ref{fig:threeQ_Cs}(a)--\ref{fig:threeQ_Cs}(i) focus on the spectra for momentum transfers in the triangular plane. Three equivalent cuts through $\mathbf{q}=(1/3,1/3,0)$, namely, along the $(1,1,0)$, $(2,-1,0)$, and $(1,-2,0)$ directions are shown, respectively, as illustrated in Fig.~\ref{fig:threeQ_Cs}(j). The calculated spectra are qualitatively similar to the observed measured neutron spectra [Figs.~\ref{fig:threeQ_Cs}(a)--\ref{fig:threeQ_Cs}(f)].
The measured integrated intensities clearly lack the characteristic double-hump structure consistently predicted by the ESWT calculations [Figs.~\ref{fig:threeQ_Cs}(g)--\ref{fig:threeQ_Cs}(i)].

\begin{table}[b]
\caption{Hamiltonian parameters obtained by fitting the ESWT calculation to the measured dispersion in \CsFe and \RbFe. Results from Refs.~\cite{Yoshizawa1980,Hayashida2019pressure} are also shown. Values are given in meV}
\label{tb:parameters}
\begin{tabular}{lcccc}
\hline \hline
\hspace{18mm} & \hspace{6mm}$J_{1}$\hspace{6mm} & \hspace{6mm}$J_{2}$\hspace{6mm} & \hspace{6mm}$J_{3}$\hspace{6mm} & \hspace{6mm}$D$\hspace{6mm} \\
\hline
\multicolumn{5}{c}{\CsFe} \\
Ref.~\cite{Yoshizawa1980} & $-0.227$ & - & $0.012$ & $2.18$ \\
Ref.~\cite{Hayashida2019pressure} & $-0.25$ & - & $0.0156$ & $2.345$ \\
This paper & $-0.268(1)$ & $0.045(2)$ & $0.016(1)$ & $2.144(4)$ \\
\multicolumn{5}{c}{\RbFe} \\
Ref.~\cite{Yoshizawa1980} & $-0.271$ & - & $0.025$ & $2.43$ \\
This paper & $-0.318(2)$ & $0.054(2)$ & $0.025(1)$ & $2.317(9)$ \\
\hline \hline
\end{tabular}
\end{table}

\begin{figure*}[tbp]
\includegraphics[scale=1]{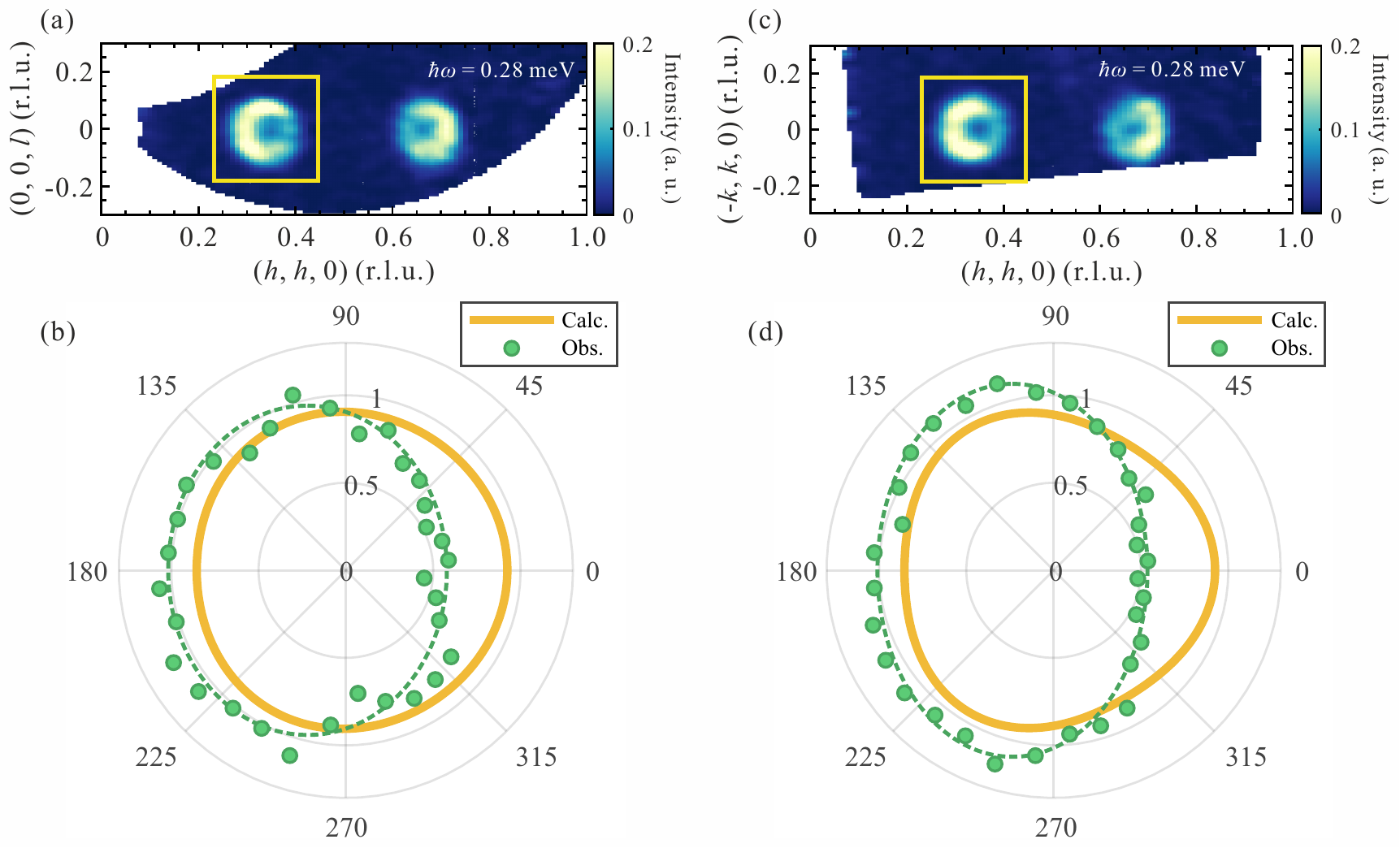}
\caption{(a) and (c) Constant-energy slices of the INS spectra integrated in $0.28\pm 0.025$~meV for \RbFe. The spectra are projected onto the (a) $(h,h,0)$-$(0,0,l)$ and (c) $(h,h,0)$-$(-k,k,0)$ planes.
(b) and (d) Corresponding polar plots of the angular intensity distributions. The data point distance from the plot center is proportional to neutron intensity integrated within narrow sectors in the $(-1,1,0)$ and $(0,0,1)$ planes, respectively. The polar angle is measured relative to the $(1,1,0)$ direction. The solid curves are the ESWT result. The dashed curves are guides to the eye.}
\label{fig:polar}
\end{figure*}

\subsection{\RbFe: magnetically ordered state}\label{sec:RbFeCl3}
In the magnetically ordered \RbFe, we observe gapless linear dispersion at magnetic Bragg peaks $\mathbf{q}=(1/3,1/3,0)$ and $(2/3,2/3,0)$ and a dispersive excitation above the gapless modes, as shown in Figs.~\ref{fig:INS_Rb}(a)--\ref{fig:INS_Rb}(d).
The observed dispersion is analyzed in the same way as \CsFe.
For data in the $ab$ plane, the fitted peak linewidths vary between the computed energy resolution of the instrument (approximately 0.08~meV at a typical 0.5-meV energy transfer for $E_i=2.99$~meV) at $(1/2,1/2,0)$ and about twice that value at $(1/3,1/3,0)$. Unlike in the case of \CsFe, where the dispersion is rather flat at low energies, for the steep dispersion in \RbFe the fitted linewidth appears to be largely due to a finite wave-vector resolution and focusing effects.
The evaluated Hamiltonian parameters are listed in Table~\ref{tb:parameters}. 
Even in the best fit, $\chi^{2}=9.8$, which is considerably worse than was obtained for the Cs-based species.
The discrepancies are quite apparent in a comparison with intensities simulated based on the fit results [Figs.~\ref{fig:INS_Rb}(e)--\ref{fig:INS_Rb}(h)]. 
The main inconsistencies are for momentum transfers in the $(h,h,0)$ direction, particularly near $(1/2,1/2,0)$. 
Here, the observed energy of the well-defined magnon peak is higher than the ESWT estimates from the spin-wave velocity, which is well defined through lower-energy data. Moreover, ESWT predicts a pair of identical branches that emanate from the $(1/3,1/3,0)$ and $(2/3,2/3,0)$ Bragg peaks, respectively, and cross at  $(1/2,1/2,0)$. In contrast, the data seem to show an acoustic branch connecting the two Bragg peaks and a separate optical mode.

These discrepancies become even more apparent if one considers cuts through the $(1/3,1/3,0)$ magnetic Bragg peak along the $(1,1,0)$, $(2,-1,0)$, and $(1,-2,0)$ directions. 
For ESWT, these directions are {\em equivalent}, as can be seen from the simulations in Figs.~\ref{fig:threeQ_Rb}(d)--\ref{fig:threeQ_Rb}(f). 
This has to do with the residual threefold degeneracy of the spin-wave branches. Each branch may have nonsymmetric structure factors and unpolarized-neutron polarization factors, but their total combined intensity, modulo the ionic magnetic form factor, has threefold symmetry around $(1/3,1/3,0)$. 
Compared with the experiment [Figs.~\ref{fig:threeQ_Rb}(a)--\ref{fig:threeQ_Rb}(c)], the relative intensities of calculated magnon branches are clearly different, as are the intensity distributions within each mode.
This observation alone signifies a {\em qualitative} failure of the ESWT calculation and/or of Hamiltonian~(\ref{eq:Hamiltonian}).

The most striking discrepancy is that the measured neutron intensities are very {\em anisotropic} near the magnetic zone center even at the lowest energies, whereas ESWT predicts an isotropic intensity distribution. 
This is borne out in Figs.~\ref{fig:polar}(a) and \ref{fig:polar}(c), which show 0.05-meV-thick constant-energy slices projected onto the $(h,h,0)$-$(0,0,l)$ and $(h,h,0)$-$(-k,k,0)$ planes.
The measured intensity distribution is crescent shaped in both slices.
The corresponding polar intensity plots are kidney shaped, with a twofold azimuthal symmetry, as shown in Figs.~\ref{fig:polar}(b) and \ref{fig:polar}(d).
In ESWT, the intensity distribution around the magnetic zone center is isotropic in the low-energy--long-wavelength limit.
As mentioned above and as can be seen in  Figs.~\ref{fig:polar}(b) and \ref{fig:polar}(d), it still retains at least a threefold symmetry at the finite energy transfer of 0.28~meV. This again clearly contradicts the observations.

\section{Discussion and concluding remarks}

ESWT based on Hamiltonian~(\ref{eq:Hamiltonian}) has its limits in describing magnetic excitations in either material. 
In the case of \RbFe, the discrepancies are substantial and qualitative in nature. 
The failure of our approach at low energies, i.e., in the long-wavelength limit, is particularly telling. 
This inconsistency cannot be resolved by including additional exchange interactions into the model. According to the Hohenberg-Brinkman first moment sum rule~\cite{Hohenberg1974} applied to an exchange spin Hamiltonian~\cite{ZaliznyakLee_MNSChapter}, the total spin-wave intensity (energy-weighted sum over all dispersion branches) is a smooth trigonometric function of the wave vector. 
Nonanalytical behavior can sometimes occur near mode crossing, where intensities are rapidly redistributed between interacting branches. 
A prime example is spin waves in a honeycomb Heisenberg ferromagnet in the vicinity of Weyl points~\cite{Shivam2017,Elliot2021}. 
The resulting crescentlike intensity patterns are superficially  similar to those observed in this paper. 
However, in \RbFe the unusual azimuthal intensity modulation occurs in acoustic spin waves, far removed from any optical branches. 
This singular behavior at long wavelengths must be a manifestation of {\em long-range} spin interactions with nonanalyticities in their Fourier transforms at small momenta. 
The most likely culprit is dipolar coupling.  
As mentioned in the Introduction, the latter is well known to be relevant, at least for \RbFe, where dipolar coupling results in incommensurate magnetic phases just below the paramagnetic transition~\cite{Shiba1982,Shiba1983,Suzuki1983dipole}.  
A good example of dipolar interactions resulting in singularities in neutron scattering intensities is the famous spectral pinch points in rare-earth-based spin ice materials~\cite{Bramwell2001Science,Bramwell2001PRL,Isakov2004,Henley2005,Fennell2009}. 
Theoretical studies also demonstrate that long-range interactions give rise to anomalous damping of magnons in the long-wavelength limit~\cite{Syromyatnikov2010,Batalov2015}.
We suggest that long-range interactions may be relevant for the spin dynamics in \CsFe and \RbFe as well. 
The next step would be analytical or numerical calculations of spin-wave spectra with long-range dipolar coupling taken into account, for a direct comparison with experiment.

\begin{acknowledgements}
We thank Dr.~K.~Yu.~Povarov and Dr.~V.~K.~Bhartiya for helpful comments on the spin-wave analysis.
This work was supported by Swiss National Science Foundation under Division II.
The neutron scattering experiment at the CNCS used resources at the Spallation Neutron Source,
a Department of Energy Office of Science User Facility operated by the Oak Ridge National Laboratory (IPTS-21713.1).
\end{acknowledgements}


%

\end{document}